\documentclass[final,10pt,journal,a4paper]{IEEEtran}

\usepackage{amsmath,amsthm,amssymb}
\usepackage{braket}
\usepackage{graphicx}
\usepackage{xcolor}

\title{Classification using a two-qubit quantum chip}
\author{Niels Neumann \\
Department of Cyber Security and Robustness, TNO \\
Anna van Buerenplein 1, 2595DA \\
The Hague, The Netherlands\\
Email: niels.neumann@tno.nl}

\begin{document}

\maketitle

\begin{abstract}
Quantum computing has great potential for advancing machine learning algorithms beyond classical reach. Even though full-fledged universal quantum computers do not exist yet, its expected benefits for machine learning can already be shown using simulators and already available quantum hardware. In this work, we focus on distance-based classification using actual early stage quantum hardware. We extend earlier work and present a non-trivial reduction of a distance-based classification algorithm that uses only two qubits. The algorithm is run on a two-qubit silicon-spin quantum computer and can be used for benchmarking it against classical devices. We show that the obtained results with the two-qubit silicon-spin quantum computer are similar to the theoretically expected results. 
\end{abstract}

\section{Introduction}
Whether we are aware of it or not, machine learning has taken a prominent role in our lives. For example, various algorithms are used to process (handwritten) text~\cite{HDVK_HandwrittenCharacterRecognitionNN_1995,KKSH_SVMforTextureClassification_2002} and speech~\cite{GMH_SpeechRecognitionDeepRecurrentNN_2013,AMJDPY_SpeechRecognitionWithCNN_2014}. Furthermore, machine learning algorithms exist to recognise patterns~\cite{Bishop_PatternRecognitionMachineLearning_2006} and also more specifically focused on recognising faces~\cite{LGAB_FaceRecognitionCNN_1997}. These are just a few of the many applications of machine learning. 

In general however, we can distinguish between three different types, being supervised, unsupervised and reinforced machine learning. In supervised machine learning, annotated data is given to the machine, which then trains a model to predict the label or annotation of unseen data. Examples of supervised machine learning are decision trees, support vector machines and neural networks. Unsupervised machine learning uses data without annotation and instead assigns the labels itself. Examples of unsupervised machine learning are clustering algorithms, such as $k$-means clustering. The third and last type is reinforced machine learning, or reinforcement learning, where a reward function is used that quantifies the `goodness' of the solution. Reinforcement learning is for instance used to train AlphaGo, the first machine that beat a human at the game of Go~\cite{Deepmind_MasteringGo_2016}. Model parameters are consequently adjusted based on the value of this reward function. As annotated data is in general expensive to gather, often a fourth type is considered: semi-supervised learning. Here a model is initially trained with a small set of annotated data, after which the rest of the learning is done unsupervised with a larger set of data without labels.

Each of these types has its own challenges. Two common challenges across the four different types are however lack of data and intractability of the training phase, meaning that training the model is too complex. The intractability is often overcome by running the algorithms on stronger (clusters of) computers. Instead, we may also opt for a completely different way of computing: doing computations using quantum computers. With quantum computers, computations are doing using quantum bits (qubits) instead of classical bits. We can thereby exploit quantum mechanical principles such as superposition of definite classical states. In contrast to classical systems, quantum systems can be considered as complex linear combinations of definite classical states. Furthermore, quantum mechanics allows for entanglement between different states, which gives correlations beyond what is classically possible. Classical information can be extracted from a quantum computer by doing measurements. However, quantum measurement are probabilistic operations, where a single output is obtained from the respective possible results, with certain corresponding probabilities. The possible output states correspond to the states present in the superposition. The original state collapses into the measured state, and hence, information present in the original state might be lost. An introduction to quantum computing is given in~\cite{NC_QuantumComputingQuantumInformation_2010}.

These quantum mechanical properties can be used to enhance classical computing and machine learning specifically~\cite{SSP_IntroductionQML_2015,NPV_MachineLearningQuantum_2019}. Computationally expensive subroutines in classical algorithms can for instance be replaced by an efficient quantum equivalent, thereby enhancing the algorithm as a whole. Examples are sampling from probability distributions~\cite{PBRB_OpportunitiesChallengesQuantumAssistedML_2018} and matrix inversion~\cite{HHL_SolvingLinearSystemsQuantum_2009}. 

Another example of where quantum computing will provide improvements over classical algorithms is classification. Specifically for distance-based classification, where a label is assigned based on a distance measure evaluated on the features of the training points and of a new test point, the complexity is $\mathcal{O}(NM)$, with $N$ the number of data points and $M$ the number of features. In \cite{SFP_DistanceBasedClassifierQuantum_2017} a quantum version of a distance-based classifier is proposed that has constant complexity $\mathcal{O}(1)$, independent of the size of the data set, while the same performance is obtained as classically. 

In this work we extend upon this distance-based classifier. We present a non-trivial reduction of the algorithm to be used on few qubit quantum computers. The proposed algorithm classifies a data point based on two classes, each of a single data point. The algorithm can be used as benchmarking algorithm for comparison with classical devices. In this work we compare the classical theoretical results with the results obtained through quantum simulation and we running the algorithm of a two-qubit silicon-spin quantum chip. In Section~\ref{sec:DistanceBasedClassifier} we briefly explain the distance-based classifier presented in~\cite{SFP_DistanceBasedClassifierQuantum_2017}. In Section~\ref{sec:TwoQubitReduction} we present a non-trivial reduction of the algorithm to a two-qubit version. The results of running the algorithm on the hardware and on the simulator are presented in Section~\ref{sec:Results}. These results are compared to the expected theoretical results. Conclusions are given in Section~\ref{sec:Conclusions}. 

\section{\label{sec:DistanceBasedClassifier}Distance-based classifier}
In this section we explain the quantum distance-based classifier proposed in~\cite{SFP_DistanceBasedClassifierQuantum_2017} and consider the technicalities when implementing and using the algorithm. This algorithm classifies a new test point, based on its distances to data points in a data set. The output of the algorithm is a bit representing the label of the test point. 

Consider a data set $\mathcal{D} = \{{\bf x}^i, y^i\}_{i=0}^{N-1}$ with data points ${\bf x}^i\in\mathbb{R}^{M}$ and label $y^i\in\{\pm 1\}$. Let ${\bf \tilde{x}} \in \mathbb{R}^{M}$ be an unlabeled data point, the goal is to assign the new label $\tilde{y}$ to this data point ${\bf \tilde{x}}$. The algorithm presented in~\cite{SFP_DistanceBasedClassifierQuantum_2017}, implements the threshold function
\begin{equation}
\label{eq:ThresholdFunction}
\tilde{y} = {\rm sgn}\left(\sum_{i=0}^{N-1} y^i \left[ 1 - \frac{1}{4N}\left\lVert {\bf\tilde{x}} - {\bf x}^i\right\rVert^2\right]\right),
\end{equation}
where ${\rm sgn}\colon\mathbb{R}\rightarrow \{-1,1\}$ is the signum function. Here $\kappa({\bf\tilde{x}}, {\bf x}) = 1 - \frac{1}{4N}\left\lVert {\bf\tilde{x}} - {\bf x}\right\rVert^2$ is the similarity function or kernel.

Without loss of generality we can assume that the data points in $\mathcal{D}$ are normalised. The data points can then be encoded in the qubits:
$${\bf x}=(x_0, \hdots,x_{M-1})^{T} \mapsto \ket{{\bf x}} = \sum_{j=0}^{M-1} x_j \ket{j},$$
with $x_j$ the $j$-th coefficient of ${\bf x}$ and $\ket{j}$ the $j$-th computational basis state. Suppose we start with the following quantum state
\begin{equation}
\label{eq:InitialQuantumState}
\ket{\mathcal{D}} = \frac{1}{\sqrt{2N}}\sum_{i=0}^{N-1} \ket{i}\left(\ket{0}\ket{{\bf \tilde{x}}} + \ket{1}\ket{{\bf x}^i}\right) \ket{y^i}.
\end{equation}
Here the first register $\ket{i}$ is an index register, indexing the data points. The second register is an ancilla qubit entangled with the new test point and the $i$-th data point. The fourth register encodes the label $y^m$. In case of only two classes, the fourth register is only a single qubit and we identify $y^m = s$ with $\ket{y^m} = \ket{(s+1)/2}$.

The algorithm is now given by a Hadamard operation on the ancilla qubit, a measurement of that qubit and a measurement of the fourth register. Due to the probabilistic nature of quantum algorithms, multiple measurement rounds should be used. The label of the test point is assigned based on the measurement of the fourth register, conditional on the first measurement giving a $0$. Results where the first measurement gives a $1$ should be neglected. 

After the Hadamard gate we are left with 
\begin{equation}
\label{eq:AfterHadamardState}
\ket{\mathcal{D}} = \frac{1}{2\sqrt{N}}\sum_{i=0}^{N-1} \ket{i}\left(\ket{0}(\ket{{\bf \tilde{x}}} + \ket{{\bf x}^i}) + \ket{1}(\ket{{\bf \tilde{x}}} - \ket{{\bf x}^i})\right) \ket{y^i}.
\end{equation}
Measuring the ancilla qubit and only continuing with the algorithm if the $\ket{0}$-state is measured, leaves us with
\begin{equation}
\label{eq:AfterHadamardState}
\ket{\mathcal{D}} = \frac{1}{2\sqrt{Np_{acc}}}\sum_{i=0}^{N-1} \ket{i}\ket{0}(\ket{{\bf \tilde{x}}} + \ket{{\bf x}^i})\ket{y^i}.
\end{equation}
Here, $p_{acc}$ is the probability of measuring $0$, given by 
\begin{equation}
p_{acc} = \frac{1}{4N}\sum_i\left\lVert {\bf\tilde{x}} + {\bf x^i}\right\rVert^2.
\label{eq:AcceptanceProbability}
\end{equation}
If instead $1$ is measured, the algorithm should be restarted. This can also be taken care of in a post-processing step. The output bit is obtained by measuring the fourth register, which encodes the labels. The probability of obtaining $\tilde{y}^m=1$ is given by
\begin{equation}
\mathbb{P}(\tilde{y}=1) = \frac{1}{4Np_{acc}}\sum_{i|y^i=1} \left\lVert {\bf \tilde{x}} + {\bf x}^i\right\rVert^2.
\label{eq:ProbilityYis1}
\end{equation}
If both classes have the same number of data points and the data points are normalized, we have
$$\frac{1}{4N}\sum_i\left\lVert {\bf\tilde{x}} + {\bf x}^i\right\rVert^2 = 1 - \frac{1}{4N}\sum_i\left\lVert {\bf\tilde{x}} - {\bf x}^i\right\rVert^2.$$
This algorithm thus implements the classifier of Eq.~\eqref{eq:ThresholdFunction}. Running multiple measurement rounds, the most likely class is obtained. 

Note that the constant complexity $\mathcal{O}(1)$ of this algorithm assumes an efficient state preparation, for instance using a quantum RAM~\cite{GLM_QuantumRAM_2008}. In~\cite{WNP_DistanceBasedClassifier_2020} implementations of the distance-based classifier are given and assumptions made in the original work are relaxed. In that same work, extensions are given towards non-balanced classes. In the next section we will reduce this algorithm to a two qubit version. 

\section{\label{sec:TwoQubitReduction}Reduction to a two-qubit version}
In this section we present a non-trivial reduction of the distance-based classifier to a two qubit version. The algorithm proposed in this section produces the same probability distribution for the measured labels as the original distance-based classifier for a given data set. In our approach, we use the same qubit for both encoding the data points as well as encoding the labels. 

For a two-qubit version of the algorithm, we consider a training set $\mathcal{D}=\{({\bf x}^0,-1), ({\bf x}^1, 1)\}$ and a test point ${\bf \tilde{x}}$, all with two features. We can encode our data points as
\begin{align*}
\ket{{\bf x}^0} & = \cos(\theta/2)\ket{0} - \sin(\theta/2)\ket{1} \\
\ket{{\bf x}^1} & = \cos(\phi/2)\ket{0} - \sin(\phi/2)\ket{1} \\
\ket{{\bf \tilde{x}}} & = \cos(\omega/2)\ket{0} - \sin(\omega/2)\ket{1},
\end{align*}
such that $R_y(\theta)\ket{0} = \ket{{\bf x}^0}$. Without loss of generality we may assume $\theta=0$. Furthermore, note that for two data points, the index register and the label register have the same value. Hence, the two can be combined. This gives as initial state
\begin{align*}
&\frac{1}{2} \ket{0}\big( \ket{0}\ket{{\bf \tilde{x}}} + \ket{1}\ket{0}\big) \\
+& \frac{1}{2} \ket{1}\big( \ket{0}\ket{{\bf \tilde{x}}} +\ket{1}\ket{{\bf x}^1}\big).
\end{align*}
The ratio of the probabilities when measuring the first register is then given by
\begin{equation}
\frac{\mathbb{P}(\ket{y^m} = \ket{0})}{\mathbb{P}(\ket{y^m} = \ket{1})} = \frac{\cos^2\left(\frac{\omega}{4}\right)}{\cos^2\left(\frac{\omega-\phi}{4}\right)},
\label{eq:RatioLabelQubit}
\end{equation}
with $\omega$ and $\phi$ depending on our data.

\begin{figure}
    \centering
    \includegraphics[width=\columnwidth]{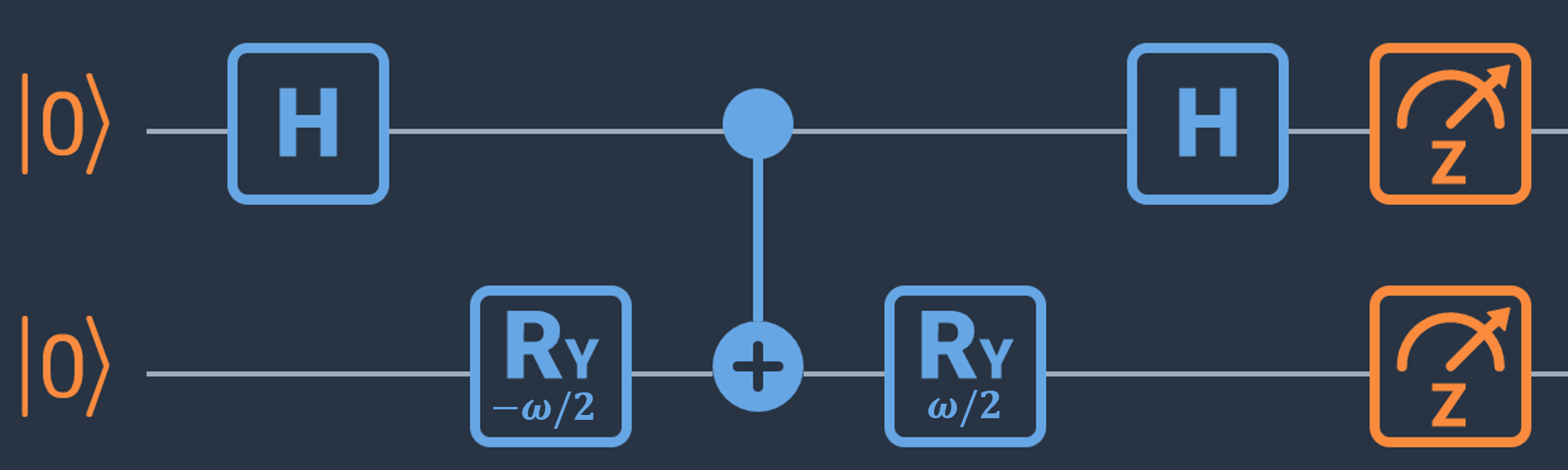}
    \caption{A two qubit classification quantum circuit. The operations are a Hadamard gate ($H$), rotations around the $Y$-axis ($R_y$), a controlled-$NOT$ operation ($CNOT$) and two measurements. The used angle depends on the data points.}
    \label{fig:QuantumCircuitTwoQubits}
\end{figure}

Now, we reduce the algorithm to only two qubits. Let $t= \cos^2\left(\frac{\omega}{4}\right) / \cos^2\left(\frac{\omega-\phi}{4}\right)$, then, we define
\begin{equation}
\omega'=4\arctan \left(\frac{1-\sqrt{t}}{1+\sqrt{t}}\right)
\label{eq:AngleTwoQubit}
\end{equation}
if $t\neq 1$. If $t=1$, both classes are equally likely and we define $\omega'=0$. We propose the quantum circuit shown in Fig.~\ref{fig:QuantumCircuitTwoQubits}, which produces the same probability distribution as the original classifier. 

The last Hadamard-gate and the measurements of both qubits shown in Fig.~\ref{fig:QuantumCircuitTwoQubits} remain the same as in the original algorithm. The other operations are state preparation. Note that state preparation gives
\begin{equation}
\frac{1}{\sqrt{2}}\left(\cos(\omega'/2)\ket{00} + \sin(\omega'/2)\ket{01} + \ket{11}\right).
\label{eq:InitialStateTwoQubits}
\end{equation}

The Hadamard-gate on the first qubit results in
\begin{align*}
&\frac{1}{2}\left(\cos(\omega'/2)\ket{00} + \cos(\omega'/2)\ket{10}\right. \\
+&\left. (1+\sin(\omega'/2))\ket{01} + (1 - \sin(\omega'/2))\ket{11}\right).
\end{align*}
When measuring the first (left-most) qubit and keeping only continuing if the $\ket{0}$-state is measured, we have
\begin{equation}
\frac{1}{2\sqrt{p_{acc}'}} \left(\cos(\omega'/2)\ket{00} + (1+\sin(\omega'/2))\ket{01}\right).
\label{eq:StateAfterMeasurementTwoQubits}
\end{equation}
Here, $p_{acc}'$ is the acceptance probability, given by
\begin{equation}
p_{acc}' = \frac{1+\sin(\omega'/2)}{2}.
\label{eq:NewAcceptanceProbability}
\end{equation}
Note that this acceptance probability differs from the one given in Eq.~\eqref{eq:AcceptanceProbability}. Also note that the two qubit classification algorithm gives the same probability distribution as the original algorithm.

\section{\label{sec:Results}Results}
In this section we present the results when running the algorithm on quantum hardware and on a quantum simulator. These results are compared to the results we should obtain theoretically. The theoretical values are obtained by determining the probability distribution from the quantum state in Eq.~\eqref{eq:StateAfterMeasurementTwoQubits} and the acceptance probability from Eq.~\eqref{eq:NewAcceptanceProbability}.

We used the Quantum Inspire platform~\cite{QuantumInspire}, developed by QuTech to obtain our results. This online platform hosts two quantum chips: a 2-qubit silicon spin chip and a 5-qubit transmon chip. Furthermore, a quantum simulator based on the QX programming language is available~\cite{KAFAB_QXprogrammingLanguage_2017}. We used the publicly available 2-qubit silicon spin chip for our results.

For the experiments we use the Iris flower dataset~\cite{Fisher_IrisFlower_1936}. This data set contains $150$ data points, equally distributed over three classes. Each data point has four features. We only consider the Setosa and Versicolor class and the first two features of the data points: the width and length of the sepal leaves. We standardize and normalize the data points and sample two data sets from them. For both data sets, we randomly select a data point from both classes and label them accordingly. For the first data set, we sample the test point from the Setosa class, for the second data set we sample the test point from the Versicolor class. The three data points of each data set can now be written as
\begin{align*}
\ket{{\bf x}^0} & = \ket{0} \\
\ket{{\bf x}^1} & = \cos(\phi/2)\ket{0} - \sin(\phi/2)\ket{1}\\
\ket{{\bf \tilde{x}}} & = \cos(\omega/2)\ket{0} - \sin(\omega/2)\ket{1}, 
\end{align*}
with appropriate angles $\phi$ and $\omega$. We identify label $0$ with the Setosa class and label $1$ with the Versicolor class. 

For the first data set we have ${\bf x}^0=(1,0)$, ${\bf x}^1 = (-0.9929, 0.1191)$ and ${\bf \tilde{x}} = (0.9939, 0,1103)$, which correspond with Iris samples $34$, $75$ and $13$, respectively. The corresponding angles are $\phi\approx-6.0445$ and $\omega\approx-0.2210$. For the second data set we randomly chose Iris samples $21$, $58$ and $82$. Hence, the data points are given by ${\bf x}^0=(1,0)$, ${\bf x}^1 = (-0.1983, 0.9802)$ and ${\bf \tilde{x}} = (0.5545, 0.8322)$. The corresponding angles are $\phi\approx-3.5407$ and $\omega\approx-1.9662$.

For both data sets, we determine $t$ and $\omega'$ and consequently run the circuit as shown in Fig.~\ref{fig:QuantumCircuitTwoQubits} on the quantum simulator and on the 2-qubit silicon spin chip. Furthermore, we compute the theoretically expected variables. The found probabilities are shown in Tab.~\ref{tab:ResultsClassification}. This table also shows the acceptance probabilities. For both the quantum hardware and the quantum simulator, we determined the probabilities based on $2048$ circuit evaluations. The shown probabilities for both labels are conditional on the ancilla qubit being in the $\ket{0}$-state.
\begin{table}
    \centering
    \caption{Shown are the results for classifying ${\bf\tilde{x}}$. Hardware and simulation results are shown as well as the theoretical values. The results hardware and simulation results are taken from $2048$ measurement rounds.}
    \begin{tabular}{l|l|c|c|c}
        & & $p_{acc}$ & $\mathbb{P}(y^m) = -1$ & $\mathbb{P}(y^m) = 1$ \\
        \hline
        & Hardware & 0.83544 & 0.7744 & 0.2256 \\
        Data set 1 & Simulation & 0.9893 & 0.9877 & 0.0123 \\
        & Theoretical & 0.9870 & 0.9870 & 0.0130 \\
        \hline
        & Hardware & 0.3755 & 0.4655 & 0.5345 \\
        Data set 2 & Simulation & 0.4863 & 0.4719 & 0.5281 \\
        & Theoretical & 0.5232 & 0.4768 & 0.5232
    \end{tabular}
    \label{tab:ResultsClassification}
\end{table}

We see that the results obtained with the simulations match the theoretical values quite well. The differences between the simulation results and the quantum hardware results follow from the decoherence of the qubits and the noise in the algorithm execution. Note that in all cases, the correct label is assinged to the test point: label $0$ for the first data set and label $1$ for the second data set. 

\section{\label{sec:Conclusions}Conclusions}
This paper presented a quantum algorithm for the practical problem of classification. The quantum distance-based classifier proposed in~\cite{SFP_DistanceBasedClassifierQuantum_2017} was non-trivially extended to a two qubit version with which a data point can be classified in one of two classes, each containing a single data point. Each data point can also represent the mean of a large set of point, thereby allowing classification for arbitrarily large data sets, after preprocessing. As the obtained results can also be computed classically, the algorithm functions as a bench-mark for near-term quantum computers. 

We found that the same probability distribution is produced as one would obtain with the original multi-qubit approach, however, less qubits are used. Due to effects such as noise and decoherence, the results for the hardware runs differ slightly from the simulated results. We tested the algorithm with two random data sets and in both cases, the correct label was assigned to the test point. The hardware used to produce these results was a two-qubit silicon-spin quantum chip, developed by QuTech and hosted publicly on the Quantum Inspire platform. 

\bibliographystyle{ieeetr}
\bibliography{paper.bib}


\end{document}